\documentstyle[aps,prl,multicol,epsf,rotate]{revtex}

\begin{document}

\title{Interlayer Transport of Quasiparticles and Cooper pairs in 
Bi$_{2}$Sr$_{2}$CaCu$_{2}$O$_{8+\delta}$ Superconductors}

\author{Yu.I. Latyshev$^*$ and T. Yamashita}

\address{Research Institute of Electrical Communication, Tohoku 
University, 2-1-1, Katahira, Aoba-ku, Sendai 980-8577, Japan}

\author{L.N. Bulaevskii$^1$, M.J. Graf$^2$, 
A.V. Balatsky$^{1,2}$, and M.P. Maley$^3$}

\address{$^1$Theory Division, $^2$Center for Materials Science,
$^3$Superconductivity and Technology Center \\
Los Alamos National Laboratory, Los Alamos, NM 87545}

\date{March 12, 1999}
\maketitle

\begin{abstract}
We study the $c$-axis transport of stacked, intrinsic junctions 
in Bi$_{2}$Sr$_{2}$CaCu$_{2}$O$_{8+\delta}$ single 
crystals, fabricated by the 
double-sided ion beam processing technique from single crystal whiskers. 
Measurements of the I-V characteristics of these samples allow us 
to obtain the temperature and voltage dependence of 
the quasiparticle $c$-axis conductivity in the superconducting state, 
the Josephson critical current, and the superconducting gap. We 
show that the BCS d-wave model in the clean limit for resonant 
impurity scattering with a significant contribution from 
coherent interlayer tunneling, describes 
satisfactorily the low temperature and low energy 
$c$-axis transport of both quasiparticles and Cooper pairs. 
\\[2ex]
{PACS numbers: 74.25.Fy, 74.50.+r, 74.72.Hs} \hfill {LA-UR: 99-1180}
\end{abstract}

\pacs{{PACS numbers: 74.25.Fy, 74.50.+r, 74.72.Hs} \hfill {LANL: LA-UR-99-1180}}

\begin{multicols}{2}

The observation of the pseudogap in the underdoped cuprate superconductors 
YBa$_2$Cu$_3$O$_{7-\delta}$, La$_{2-x}$Sr$_x$Cu O$_{4+\delta}$ 
and Bi$_{2}$Sr$_{2}$CaCu$_{2}$O$_{8+\delta}$ 
(Bi-2212) is indicative of the breakdown of the Fermi-liquid theory in 
these systems \cite{nonfermi}. 
The situation remains unclear in the overdoped regime. 
On the other hand, the superconducting state is usually discussed in the 
BCS d-wave pairing model, which is based on the Fermi liquid picture. 
Such an approach may be limited because 
(a) the properties of the normal state 
determine the mechanism of superconductivity, and (b) the ratio 
$2\Delta_0/T_c$ is well above the BCS ratio for d-wave pairing and is 
strongly doping dependent. 
Specifically, the BCS approach may fail  
in describing the properties of the superconducting state that are 
directly related to the quasiparticles, while the electrodynamics, 
based on supercurrents (macroscopic quantum phenomena), 
is almost insensitive to the pairing mechanism. 

The interlayer currents of both quasiparticles and Cooper pairs may be studied 
in highly anisotropic Bi-2212 crystals with Josephson interlayer coupling 
by measuring the I-V characteristic of the $c$-axis current. 
Such measurements provide information on the voltage and 
temperature dependence of the quasiparticle $c$-axis current, 
the energy gap and the Josephson interlayer current. 
These data allow us to check the validity of 
the BCS d-wave model and determine the degree of the coherence of the 
interlayer tunneling. The question of coherence in 
both the normal and superconducting state is the focus of numerous 
theoretical and 
experimental studies, see for example \cite{and,wata,tana,kirtley,klemm}. 
Recently, Tanabe {\it et al.} \cite{tana} 
measured the quasiparticle $c$-axis transport in the superconducting state 
of Bi-2212 crystals 
and concluded that their data support the d-wave pairing scenario. 
However, their results for the 
quasiparticle current are insufficient 
to determine the nature of the interlayer transport 
and the effect of intralayer scattering on this transport. 

Our measurements of I-V characteristics have been performed on high quality,  
stacked, intrinsic mesa junctions, fabricated from 
perfect single crystal Bi-2212 
whiskers by double-sided focused ion beam (FIB) processing \cite{lat}. 
For the fabrication we used the conventional FIB machine of Seiko Instruments 
Corp., SMI 9800 (SP) with Ga$^+$-ion beam. The details of the fabrication are 
described in Ref.~\onlinecite{lat2}. 
Here we note only that the method allows us to 
fabricate the mesa junctions with the in-plane size down to 0.5 $\mu$m without 
degradation of $T_c$. 
We studied 5 junctions with in-plane areas ranging from 6 $\mu$m$^2$ down 
to 0.3 $\mu$m$^2$, and the 2D array of 6$\times $6 stacks with area 
0.5 $\mu$m$^2$ each (see Table~I). 
The number of intrinsic junctions, $N$, 
in the stack  was typically about 50. The four leads were 
attached outside of the junction area, see  Fig.~1(a). The contact Au-pads 
were ablated and annealed before the FIB processing to avoid the 
diffusion of the Ga-ions into the junction body. 
Fig.~1(b) shows the I-V characteristic of sample \#2. 
The fully superconducting overlap geometry of the stack let us 
avoid the effects of quasiparticle injection on the tunneling 
characteristics, 
usually occurring in junctions of the mesa type with a 
normal metal top electrode\cite{tana,yurg}. 
We also substantially reduced the 
effects of self-heating in our submicron mesa junctions. Self-heating 
manifests itself in the form of an S-shaped I-V curve near 
the gap voltage $V_g$ \cite{lat3}. 
As a result, for small junctions we found the high quality 
tunneling characteristics free of the nonequilibrium effects mentioned 
above [Fig.~1(c)].
The measured temperature dependence of the $c$-axis resistivity 
of the stacks (at DC currents $\approx$1 $\mu$A) was typical for 
slightly overdoped Bi-2212 crystals with  $T_c\approx 77$ K \cite{wata}.

The critical current was determined from the I-V characteristics as  
the current of switching from the superconducting to the resistive state, 
averaged over the stack. The variation of the critical current along 
the stack is not large (usually within 15\%), indicating a good uniformity 
of our structures. 
The $c$-axis critical current density $J_c$ for the junctions with 
in-plane areas  $S > 2$ $\mu$m$^2$ was typically 
600 A/cm$^2$ at $T=4.2$ K (see Ref.~\onlinecite{lat3}), and is consistent 
with the data reported by other groups \cite{tana}. 
For bigger stacks the dependence of $J_c$ on magnetic fields parallel to the 
layers demonstrates well resolved Fraunhofer diffraction patterns \cite{lat4}, 
which prove the presence of the intrinsic DC Josephson effect in our 
stacks. $J_c$ is depressed for submicron junctions (see Table~I),
presumably due to the Coulomb blockade  effect \cite{lat3}.


\begin{figure}
\noindent
\begin{minipage}{0.48\textwidth}
\epsfysize=3.0in
\epsfxsize=0.6\hsize
\centerline{ \epsfbox{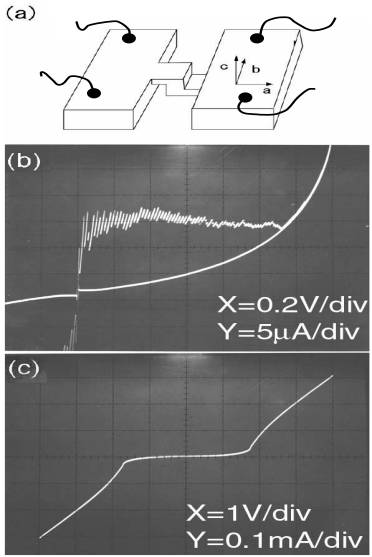} }
\vspace*{6pt}
\caption{Schematic view of the setup (a) 
and the I-V characteristics of the Bi-2212 stacks at $T=4.2$ K: 
(b) enlarged I-V scale of sample \#2; 
(c) extended I-V scale of sample \#4. 
}
\end{minipage}
\end{figure}

The superconducting gap voltage of the stack, $V_g$, was determined 
from the I-V characteristics as the voltage at the maximum of $dI/dV$. 
The gap of the intrinsic junction, 
$2\Delta_0 \approx eV_g/N$, reaches values  as high as 50 meV, 
see Table~I and Ref.~\onlinecite{lat3}.
Note that this relation would be exact for a stack of conventional
s-wave junctions.  This result is consistent  with the value recently 
found by surface tunneling measurements for slightly 
overdoped Bi-2212 single crystals \cite{miy}. 

The multibranched structure, which is clearly seen in Fig.~1(b), 
corresponds to subsequent transitions of the intrinsic junctions 
into the resistive state for increasing voltage \cite{kleiner}. 
At voltages $V>V_g$ all junctions are resistive. 
In the down-sweep of voltage, starting from $V>V_g$,  
the I-V curve  is observed in the all-junctions 
resistive state. Here only quasiparticles contribute to the 
$c$-axis DC transport. The ohmic resistance, $R_n$, at $V > V_g$ is 
well defined [see Fig.~1(c)].
This resistance is nearly temperature independent (Fig.~2) and corresponds 
to the conductivity $\sigma_n(V > V_g)\approx80$ (k$\Omega$ cm)$^{-1}$ 
for energies well above the pseudogap and the superconducting gap.

The I-V curve of the all-junctions resistive state 
[lower curve in Fig.~1(b)] at low voltages fits the dependence
\begin{equation}\label{IV_eq}
I_q(V)=\sigma_q(0) \frac{S}{s}
\left( v+\frac{b}{3}v^3 \right), \ \ \ v=V/N,
\end{equation}
where $v\leq 10$ mV, as seen in Fig.~3, 
and $s$ is the spacing between intrinsic superconducting 
layers (15.6 \AA). Values for $\sigma_q(0)$, $b$ and $S$ are given in Table~I. 
For the quasiparticle differential conductivity, 
$\sigma_q(v,T) \equiv s^{-1}\partial J_q/\partial v$, at $v\rightarrow 0$ 
we find for $T\leq T^*\approx 30$ K:
\begin{equation}
\sigma_q(T)=\sigma_q(0) \big( 1 + c T^2 \big),
\end{equation}
with $c\approx 9.6\cdot 10^{-4}$ K$^{-2}$ and $6.4\cdot 10^{-4}$ K$^{-2}$ 
for samples \#2 and 
\#3, respectively, see insert in Fig.~2. We estimate the accuracy 
for extracting $b$ and $c$ to be within $\approx$ 30\%. 

Previously, the quasiparticle conductivity 
in the superconducting state was obtained from $c$-axis conductivity 
measurements in high magnetic 
fields, which suppress the contribution of Josephson current to the 
$c$-axis transport. By this way, in magnetic fields up to 18 T, 
$\sigma_q(T)$ was obtained at temperatures above 50 K and 
found to increase with temperature \cite{cho}. 
These results are in quantitative 
agreement with those shown in Fig.~2. We note that 
data for sample \#7, taken from Ref.~\onlinecite{tana}, are also 
in quantitative agreement with ours, though the sample may be 
slightly different.


\begin{figure}
\noindent
\begin{minipage}{0.48\textwidth}
\epsfysize=0.95\hsize
\centerline{ \rotate[r]{\epsfbox{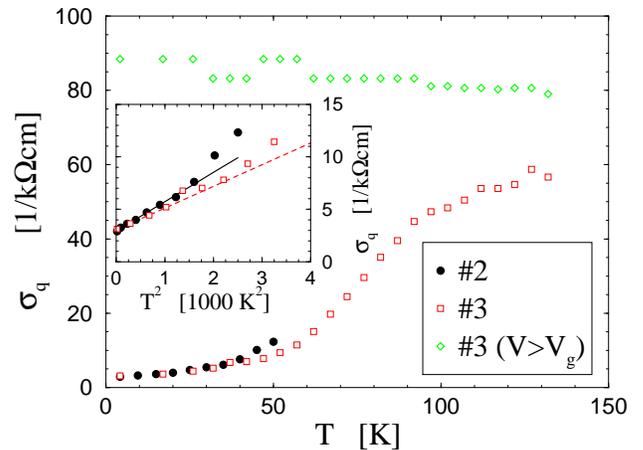}} }
\caption{Quasiparticle dynamic conductivity vs. $T$ for voltages 
$v> V_g/N \approx 2\Delta_0/e$ and $v=V/N\rightarrow 0$, as extracted from the I-V 
characteristics of samples \#2 and \#3. Inset: $\sigma_q$ vs. $T^2$ 
at voltage $v\rightarrow 0$.  Lines are fits for $T^2 < 1000\,{\rm K}^2$}
\end{minipage}
\end{figure}

Our results differ remarkably from the tunneling 
characteristics of junctions 
between conventional superconductors in two aspects. 
First, the 
value $\sigma_q(T)$ remains nonzero as $T\rightarrow 0$. 
As was mentioned above, Tanabe and coworkers also observed Ohm's law in the 
all-junctions resistive state 
at low temperatures, but considered it as being of extrinsic origin. 
Second, $J_c(T)$ at $T\rightarrow 0$ 
is substantially reduced (by a factor $\approx 30$) 
in comparison with the value given by the Ambegaokar-Baratoff relation, 
$J_c^{(AB)}(0)=\pi\sigma_n\Delta_0/(2 e s)$.
Instead, we find at $T\rightarrow 0$ for stacks with large areas  that
the relation 
\begin{equation}
J_c(0)\approx \frac{ \pi\sigma_q(0)\Delta_0 }{e s}
\label{nr}
\end{equation}
holds.
This result is expected when coherent tunneling is dominating
the $c$-axis transport. 
We obtain the ratios $e s J_c(0)/[\pi\sigma_q(0)\Delta_0]\approx 1.2$ 
and 1.5 for samples \#1 and \#2, respectively, and $\approx 0.74$ and 0.8 
for samples \#3 and \#5, 
if we take into account that their effective 
critical current is reduced in comparison with the Josephson critical current, 
due to the Coulomb blockade, and 
take a typical $J_c(0)=600$ A/cm$^2$ as for large area stacks. 


\begin{figure}
\noindent
\begin{minipage}{0.48\textwidth}
\epsfysize=0.95\hsize
\centerline{ \rotate[r]{\epsfbox{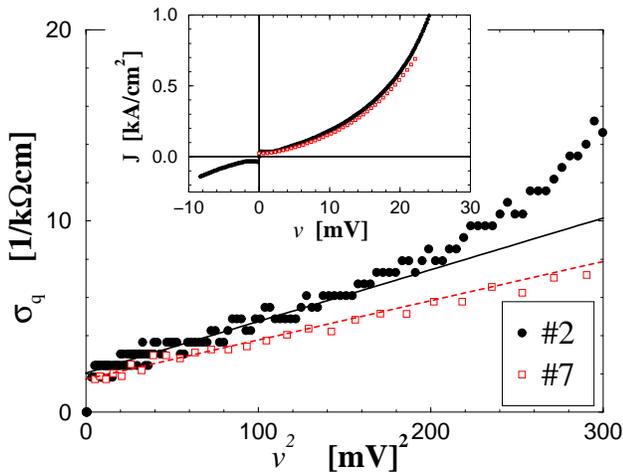}} }
\caption{The quasiparticle differential conductivity vs. $v^2=V^2/N^2$ 
at $T=4.2$ K as extracted from the I-V characteristics of sample \#2 
and \#7 (Fig.~4(c) in Ref.~[4]). Lines are fits for $v < 10$ mV.
Inset: Corresponding $J$-$v$ curves.}
\end{minipage}
\end{figure}

We analyze these experimental data in the framework of the BCS 
d-wave pairing model inside the layers, considering the 
general form of the tunneling Hamiltonian: 
\begin{equation}
{\cal H}=\sum_{n}\int \!d{\bf r}
\big[ 
t_n({\bf r})\psi_{n\sigma}^+({\bf r})
\psi_{n+1,\sigma}({\bf r}) + h.c.
\big]
+{\cal H}_{\parallel,n}.
\label{1}
\end{equation}
Here ${\cal H}_{\parallel,n}$
is the BCS Hamiltonian for d-wave pairing with isotropic intralayer 
scattering inside layer $n$, described by the bare scattering rate, $\nu_0$,
of electrons with defects. 
The superconducting gap is expressed as 
$\Delta(\varphi)=\Delta_0\cos 2\varphi$, where $\varphi$ is the angle of the 
momentum on the two-dimensional cylindrical Fermi surface. 
Further, $t_n({\bf r})$ is the random, isotropic interlayer transfer integral 
which depends on the in-plane coordinate ${\bf r}$ due to crystal 
imperfections. The correlation function 
$K({\bf r})=\langle t_n({\bf r})t_n(0)\rangle$ in the Fourier representation 
is $K({\bf q})=t_{\perp}^2[a\delta({\bf q})+(1-a)g({\bf q})]$, where 
$at_{\perp}^2=\langle t({\bf r})\rangle^2$ and 
$(1-a)t_{\perp}^2=\langle t^2({\bf r})\rangle-\langle t({\bf r})\rangle^2$. 
The weight for in-plane momentum conserving (coherent) tunneling is $a$, 
while that for incoherent tunneling is $(1-a)$. Incoherent tunneling 
is described by the normalized function 
$g({\bf q})$ with the characteristic momentum transfer $q_0$, {\it e.g.},
with a Gaussian distribution $g({\bf q})=(\pi/q_0^2)\exp(-q^2/4q_0^2)$. 
In the following,
we will discuss the case of a strongly incoherent part, when $v_Fq_0$ is of 
order $\epsilon_F$. Here $v_F$ and $\epsilon_F$ are the Fermi velocity and 
energy, respectively. 
For small transfer integral we calculate the quasiparticle 
interlayer current density, using perturbation theory with respect 
to $t_n({\bf r})$, 
\begin{eqnarray}
&&J_q(v)=\frac{es}{\pi^3\hbar}
\int_{-\infty}^{+\infty} \!d\omega\int d{\bf k}\int d{\bf q} \, K({\bf q})
\nonumber \\
&& \qquad\times
\big[ f(\omega+ev)-f(\omega) \big]
A(\omega+ev,{\bf k}+{\bf q})A(\omega,{\bf k}), 
\label{cur}
\end{eqnarray}
where $f(\omega)$ is the Fermi distribution function 
and $A(\omega, {\bf k})$ 
is the spectral density of the Green function. 
Scattering by impurities 
leads to the formation of gapless  states, $A(\omega,{\bf k})\approx 
\gamma/[(\omega-E_{\bf k})^2+\gamma^2]$ at low 
$\omega,E_{\bf k}\lesssim\gamma$. Here 
$E_{\bf k}=[\xi_{\bf k}^2+\Delta^2(\varphi)]^{1/2}$ is the quasiparticle energy 
and $\gamma$ is the impurity bandwidth
of quasiparticles \cite{lee,bra,pp,graf1,graf}.  
The quasiparticle current at $T,ev\ll\Delta_0$ comes mainly from 
regions near the gap nodes on the Fermi surface. 
The angle dependent quasiparticle density 
is sharply peaked  near the nodes at angles 
$\varphi_g\pm\varphi_0/2$, $\varphi_g=\pm \pi/4, \pm 3\pi/4$, with
$\varphi_0 \sim \gamma/\Delta_0$. 
Impurity scattering results in a nonzero 
density of states at zero energy, $N(0)\varphi_0$.  
For the coherent part this leads to a universal 
quasiparticle interlayer conductivity 
$\propto aN(0)\varphi_0/\gamma\propto aN(0)/\Delta_0$ 
at $T\rightarrow 0$. The combined parts of coherent and incoherent 
conductivities give at $v\rightarrow 0$:
\begin{equation}
\sigma_q(0)\approx\frac{2e^2t_{\perp}^2N(0)s}{\pi\hbar\Delta_0}\left[a+(1-a)
C_1\frac{\gamma}{\epsilon_F}\right], 
\label{ch}
\end{equation}
where $N(0)$ is the 2D density of states per spin at the Fermi level.   
Here and in the following $C_i$, ($i=1,2,3$), 
is a numerical coefficient of order unity. 
For the coherent part this regime 
is valid at temperatures  $T\ll T^*$, where the crossover temperature 
$T^*\approx \gamma \sim (\hbar\nu_0\Delta_0)^{1/2}$ 
for strong scattering, and $T^*\approx 4\Delta_0\exp(-\pi\Delta_0/\nu_0)$ 
in the limit of weak scattering. Finite $T$ corrections to $\sigma_q$
at $T\lesssim T^*$ are quadratic in temperature \cite{graf1}:
\begin{eqnarray}\label{c} 
\sigma_q(T) & \approx & \sigma_q(0) +
\\ &&
\frac{e^2t_{\perp}^2N(0)\pi s}{6\hbar\Delta_0}
\left(\frac{T}{\gamma}\right)^2\left[a+(1-a)C_2\frac{\gamma}{\epsilon_F}
\right]. 
\nonumber
\end{eqnarray}
Using the results of Ref.~\onlinecite{graf1}, we obtain at 
$T\to 0$ and for voltages $ev\lesssim\gamma$
for the coefficient $b$ in Eq.~(\ref{IV_eq})
\begin{equation}
b\approx \frac{1}{8\gamma^2}\left[a+(1-a)C_3\frac{\gamma^2}
{\Delta_0\epsilon_F}\right].
\label{b}
\end{equation}
The expression 
for the interlayer 
Josephson critical current density, $J_c$, 
is similar to Eq.~(\ref{cur}), but with the $A$'s replaced by the
anomalous Gor'kov functions at $v=0$. 
For the critical current density at $T=0$ and in the clean limit, 
$\hbar\nu_0\ll\Delta_0$, we obtain \cite{bul}
\begin{equation}
J_c(0)\approx \frac{2et_{\perp}^2N(0)}{\hbar}\left[a+(1-a)C\frac{\Delta_0}{
\epsilon_F}\right],
\label{crit}
\end{equation}
where $C\lesssim 1$ is a numerical coefficient, which depends on the form of 
the function $g({\bf q})$. We neglected the effect of the 
pseudogap (if any) on $J_c$ at low $T$.

We see that the contribution from incoherent tunneling to the quantities
$J_c(0)$ and $\sigma_q(0)$, and the parameters $c$ and $b$, is negligible, 
if $a \gg C\Delta_0/\epsilon_F$. Assuming that this is 
the case, we obtain the universal relation (\ref{nr}) and the ratio 
$c/b=2\pi^2/3$. Our data obey relation (\ref{nr}) as was mentioned above. 
We obtain $c/b\approx 6.9$ and 2.2 for samples \#2 and \#3, respectively. 
These numbers are in agreement with the theoretical prediction
within the accuracy of our estimates. 
 From the value of $c$ we estimate $\gamma\approx 3$ meV. 
For resonant scattering this estimate agrees well with 
the crossover temperature $T^* \approx 30$ K,
while in the Born limit $T^*$ would be very much 
smaller. Thus we discard the Born limit.  
The value of $\gamma\approx 3$ meV leads to the estimate 
$\hbar\nu_0\approx 0.4$ meV, confirming our assumption 
of the clean limit for superconductivity inside the layers. 

Thus, we obtain a self-consistent description for our 
$c$-axis transport characteristics, {\it i.e.}, for
$\sigma_q(v,T)$ and $J_c(0)$   
at low temperatures and low voltages, 
${\rm max}(T,ev)\lesssim \gamma$, 
assuming a significant contribution of 
coherent tunneling to these quantities. Note that the weight 
for coherent tunneling, $a$, should be bigger 
than $C\Delta_0/\epsilon_F\approx 0.05-0.1$, but 
may still be much smaller than unity in our model for 
the low energy behavior of the $c$-axis transport. 
It is noteworthy that $\sigma_n$, as well as  
$\sigma_q(T,v)$ at $T>\gamma$ or $ev>\gamma$, are 
related to the high energy regime and thus remain outside of 
our description of the $c$-axis transport.

We conclude that the BCS d-wave pairing model in the clean limit 
with {\it resonant} intralayer scattering and 
significant contribution of {\it coherent} interlayer 
tunneling provides a satisfactory and consistent
description of the experimental data for 
the low energy and low temperature 
interlayer transport in the superconducting state of Bi-2212 crystals 
for both quasiparticles and Cooper pairs. 
In spite of the fact that the 
normal state properties deviate from the 
Fermi-liquid behavior, we find that our data for interlayer transport 
are consistent with the interpretation that 
superconductivity restores the 
Fermi-liquid behavior of quasiparticles, at least at low temperatures.
This is in agreement with the formation of a sharp quasiparticle peak in 
the density of states 
in the superconducting state in ARPES measurements \cite{ding}. 

We thank A.M. Nikitina for providing us with Bi-2212 
single crystal whiskers, S.-J. Kim and V. Pavlenko for technical assistance, 
and M. Suzuki for sharing his data \cite{tana}. 
This work was supported  by the Los Alamos National Laboratory under the
auspices of the U.S. Department of Energy,  
CREST, the Japan Science and Technology Corporation, and 
the Russian State Program on HTS under grant No.~99016.


\begin{table}
\noindent
\begin{minipage}{0.47\textwidth}
\caption{Parameters of the stacked Bi-2212 mesa junctions.
Data for sample \#7 are from Ref.~[4].}
\begin{tabular}{ccccccccc}
No.& $S$ & $N$ & $T_c$	& $J_c$  & $V_g/N$ &  $\sigma_q(0)$ & b \\
  & ($\mu$m$^2$)&  &  (K) & (A/cm$^2$) & (mV) & (k$\Omega$~cm)$^{-1}$ 
  & (meV)$^{-2}$ \\
\hline
\#1 & 6.0    &  69	& 76 	&  600  	&   16 &  3.0 & - \\
\#2 & 2.0    &  65	& 76 	&  600  	&   20 &  2.0 & 0.014 \\
\#3 & 1.5    &  50	& 78 	&  400  	&   22 &  3.7 & 0.029 \\
\#4 & 0.6    &  34	& 76 	&  40   	&   50 &   -  & - \\
\#5 & 0.3    &  50	& 78 	&  23   	&   44 &  1.7 & 0.008 \\
\#6 & 36$\times$ 0.5& 50& 76 	&  72   	&    - &  2.5 & - \\
\#7 & 400 &  40 	& $\sim$80 & 250-500 	&   20 &  1.7 & 0.012 
\end{tabular}
\end{minipage}
\end{table}

\end{multicols}

\end{document}